\documentclass[preprint,12pt]{aastex}
\usepackage{epsfig}
\usepackage{emulateapj5}
\usepackage{apjfonts}

\newcommand{\chandra}{{\it Chandra}}

\newcommand{\rosat}{{\it ROSAT}}

\newcommand{\xmm}{{\it XMM-Newton}}

\newcommand{\lum}{\thinspace\hbox{$\hbox{erg}\thinspace\hbox{s}^{-1}$}}

\newcommand{\sss}{XMMU\,J005510.7-373855}
\newcommand{\rd}{Di\thinspace Stefano}

\makeatother

\begin{document}

\def\spose#1{\hbox to 0pt{#1\hss}}
\def\laeq{\mathrel{\spose{\lower 3pt\hbox{$\mathchar"218$}}
     \raise 2.0pt\hbox{$\mathchar"13C$}}}
\def\gaeq{\mathrel{\spose{\lower 3pt\hbox{$\mathchar"218$}}
     \raise 2.0pt\hbox{$\mathchar"13E$}}}

\title{Supersoft X-ray Sources in M31: 
I. A {\it Chandra} Survey and an Extension to Quasisoft Sources}

\author{R.~Di\,Stefano\altaffilmark{1,2}, A.~K.~H.~Kong\altaffilmark{1},
J. Greiner\altaffilmark{3}, F.~A.~Primini\altaffilmark{1},
M.~R. Garcia\altaffilmark{1}, P.~Barmby\altaffilmark{1}, 
P.~Massey\altaffilmark{4}, P.~W.~Hodge\altaffilmark{5}, 
B.~F~. Williams\altaffilmark{1}, S.~S.~ Murray\altaffilmark{1}, 
S.~Curry\altaffilmark{2}, T.~A.~Russo\altaffilmark{2}}

\altaffiltext{1}{Harvard-Smithsonian Center for Astrophysics, 60
Garden Street, Cambridge, MA 02138; rdistefano@cfa.harvard.edu}
\altaffiltext{2}{Department of Physics and Astronomy, Tufts
University, Medford, MA 02155}
\altaffiltext{3}{
Max-Planck-Institut f{\" u}r extraterrestrische Physik, Giessenbachstrasse Postfach 1312,
D-85748 Garching, Germany
}
\altaffiltext{4}{
Lowell Observatory, 1400 West Mars Hill Road, Flagstaff, AZ 86001 
}
\altaffiltext{5}{
 Astronomy Department, University of Washington, Box 351580, Seattle, WA 98195-1580
}

\begin{abstract}
We report on very soft X-ray sources (VSSs) in M31.
In a survey which was most sensitive to soft sources in four $8' \times 8'$
regions covered by {\it Chandra}'s ACIS-S S3 CCD,
we find $33$ VSSs that appear to belong to M31.
Fifteen VSSs have spectral characteristics
mirroring the supersoft X-ray sources studied in the 
Magellanic Cloud and Milky Way ($k\, T_{eff} \leq 100$ eV); 
we therefore call these ``classical'' supersoft sources, or simply 
supersoft sources (SSSs).
Eighteen VSSs  may either have small ($< 10\%$)
hard components, or slightly higher 
effective temperatures (but still $< 350$ eV).
We refer to these VSSs as quasisoft sources (QSSs).
While hot white dwarf models may apply to SSSs, the effective 
temperatures of QSSs are too high, unless, e.g., the 
radiation emanates from only a small portion of
surface.  Two of the SSSs 
were first detected and identified as such through {\it ROSAT} observations.  
One SSS and  one QSS may be identified with symbiotics, and $2$ SSSs with
supernova remnants. Both SSSs and QSSs in the disk 
are found near star-forming regions,
possibly indicating that they are young. 
VSSs
 in the outer disk and halo
are likely to be old systems; in these regions, there are more
QSSs than SSSs, which is opposite to what is found
in fields closer to the galaxy center.     
The largest density of bright VSSs is in the bulge; some of the bulge
sources
are close enough to the nucleus to be remnants of the tidal disruption of  
a giant by the massive central black hole. 
By using {\it Chandra} data in combination with {\it ROSAT}
and {\it XMM} observations, we find most VSSs to be highly variable, 
fading from 
or brightening toward detectability on time scales of months.
There is evidence for 
VSSs with low luminosities ($\sim 10^{36}$ erg s$^{-1}$).

\end{abstract}


\section{Introduction} 

\subsection{Observations of Supersoft Sources} 

	Very soft X-ray sources, with little or no apparent emission
above $1$ keV, were discovered in the Magellanic Clouds by the 
{\it Einstein} X-Ray Observatory (Long, Helfand, \& Grabelsky 1981;
Seward \& Mitchell 1981). The {\it ROSAT} All-Sky Survey detected
approximately one dozen 
such soft sources in the Magellanic Clouds, while $15$ soft sources
were discovered during  
{\it ROSAT's} 
initial survey of M31 (Supper et al.\,1997). Later, Kahabka (1999) suggested 
that, given the effects of foreground and intrinsic absorption,  
$34$ M31 X-ray sources
could have similarly soft spectra. 
Astronomers coined the term
``luminous supersoft X-ray source'' (SSS), and a new class of
X-ray sources was born. With no definite physical model,
sources were afforded membership based simply on their observable
characteristics.  
In general, there was little or no emission above $1-1.5$ keV;
blackbody spectral fits  yielded values of $k\, T$ between $30$ and $100$ eV;
the associated luminosities were $10^{37}-10^{39}$ erg s$^{-1}$.

Because the soft
X-radiation they emit is readily absorbed by the ISM, any X-ray 
census of SSSs is
highly incomplete.  
Based on the data and models of the gas distribution in M31 and in the Milky Way,
Di\thinspace Stefano \& Rappaport (1994) concluded that each 
of these $2$ galaxies likely houses a 
population of $\sim 1000$ SSSs with $L\geq 10^{37}$ erg s$^{-1}$ and
$k\, T \geq 30$ eV.  
Lower luminosity, lower temperature sources, which are
less likely to be detected, had not been discovered and were not
included in their simulations. Since then however, it has been established
that some CVs may turn on as SSSs (Greiner \& \rd\ 1999; 
Greiner et al.\ 1999; Patterson et al.\ 1998),
exhibiting temperatures ($<30$ eV) and luminosities 
($10^{35}-10^{36}$erg$ s^{-1}$)
lower than those of the original set of 
Milky Way and Magellanic Cloud SSSs. Bright CVs provide a much larger pool (perhaps $\sim 10^4$ in a galaxy like M31)
of potential SSSs. 
 
\subsection{Models} 

Two circumstances pointed to white dwarf models as promising
explanations for SSSs.
First, the effective radii for the SSSs discovered in the Galaxy and
Magellanic Clouds
are comparable to those of white dwarfs (WDs).
Second, some hot WDs and pre-WDs have been observed as SSSs
(see Greiner 2000),
including several recent novae, symbiotic systems,
 and a planetary nebula (PN).
More than half
of all SSSs with optical IDs do not however, seem to be examples
of systems of known types. It is the mysterious nature of these
other systems that has excited much of the interest in SSSs.
A promising model
is one in which matter from a Roche-lobe-filling
companion
accretes onto the WD at rates so high ($\sim 10^{-7} M_\odot$ yr$^{-1}$)
that it can undergo quasi-steady
nuclear burning (van den Heuvel et al. 1992;
Rappaport, \rd\ \& Smith 1994; Hachisu, Kato, \& Nomoto 1996,
\rd\ \& Nelson 1996).
Because matter that is burned can be retained by the WD,
some SSS binaries may be
progenitors of Type Ia
supernovae.

There is indirect evidence in favor of nuclear-burning WD
 models for several SSSs.
There is, e.g., rough agreement between optical and UV data and predictions
based on a reprocessing-dominated disk (Popham \&
Di\thinspace Stefano 1996).
In addition, the predicted velocities of bipolar outflows (jets) are near
the escape velocity
of a WD (Southwell et al.\, 1996, Parmar et al.\, 1997).
Presently, however, there is little direct evidence that the models
are correct. (See \rd\ \& Kong 2003a for a more complete
discussion.)  
Note that the highest white dwarf effective temperature 
possible in principle  is $\sim 150$ eV; this is the value associated
with the smallest possible white dwarf radiating at the Eddington limit. 
Since, however, the photosphere is expected to lie above the surface,
the maximum expected values of $k\, T$ are closer to $80-100$ eV. 

Because the phenomenological definition of SSSs is so broad,
it is in fact likely that the class includes objects of several types. 
Indeed,
any object more compact than a WD could certainly act as
an SSS. Neutron star models have been considered 
(Greiner et al.\, 1991, Kylafis \& Xilouris 1993). Although
neutron star luminosities and 
photospheric radii can be large enough to support SSS behavior,
there is no well-understood reason why such a large
photosphere would be
preferred.     
Intermediate-mass black holes (BHs) are, on the other hand, expected to 
emit as SSSs, at least if the accretor mass and luminosity
is in the appropriate range. (See \S 8). 

We also note that the stripped cores of
giant stars that have been tidally disrupted by massive
black holes are expected to
 appear as SSSs for times
ranging from $10^3$ to $10^6$ years (\rd\ et al. 2001).
Several such stripped cores could be present within
$\sim 1$ kpc of the nuclei of galaxies harboring high-mass black holes.

\subsection{SSSs in External Galaxies}

To avoid the worst effects of Galactic 
absorption, our best hope to study galactic populations of SSSs 
is to search for them in other galaxies. 
The advent of {\it Chandra}, with its good low-energy sensitivity, its superb
angular resolution, and its low background, will 
significantly increase the numbers of known sources and extend our
knowledge of the class.

Among external galaxies, M31
can play a unique role, simply because of its proximity.
First, the population of low-$L$ sources can best be studied in
M31. Second, M31 is the only galaxy in which we can hope to identify
optical counterparts to a large fraction of SSSs. This step
will allow us to get a better understanding of the natures of
SSSs, which is our primary goal.
In this paper we report on the SSSs detected by {\it Chandra} in M31.

\subsection{Quasisoft Sources}

To develop a set of criteria that would identify M31's SSSs in a
uniform and systematic way, we first studied a large number of sources
individually. We found that, for many 
(roughly $5-15\%$ of all sources), 
the majority of photons had energies below
$1.1$ keV,
as expected for SSSs. 
But we also found that a large fraction of these soft sources
produced photons of higher energy. We therefore designed our selection 
procedure to be hierarchical. Sources selected as soft by the first $2$
steps are almost certain to have intrinsic 
spectral properties (with $k\, T < 100$ eV)
similar to those that had been observed for SSSs
in the Galaxy and Magellanic Clouds and also in {\it ROSAT's} survey
of M31. Sources selected by higher steps could correspond to
the hottest known SSSs, especially if they lie behind large columns of gas;
but they could also correspond to sources that, though soft,
are intrinsically harder than the local SSSs. We therefore refer to
the sources selected by the higher steps as quasisoft sources 
(QSSs). 
We use the phrase ``very soft source" (VSS) to refer
to any source that is either supersoft or quasisoft.

We have now had the opportunity to
test our selection procedure on {\it Chandra} and some {\it XMM-Newton} 
data from roughly $20$ galaxies, many with
 less line-of-sight
absorption than is found along the direction to M31.
These galaxies include ellipticals and spirals 
viewed at a variety of inclination angles.  
(M101, M83, M51, NGC 4697 [\rd\ \& Kong 2003b, c];
M104 [\rd\ et al. 2003]; NGC 4472 [Friedman et al. 2003, \rd\ et al. 2004a];
to {\it Chandra} data from approximately $10$ other galaxies
\rd\ et al. 2004b; 
and to {\it XMM} data from
NGC 300 (Kong \& \rd\ 2003) and from the halo of M31
(\rd\ \& Kong 2003d).
We have identified hundreds of VSSs, roughly half of them QSSs.
A small group of these (including r1-9 in this paper; see Figure 2),
provide enough photons for spectral fits.
Some QSSs are well fit by thermal or disk blackbody models with
$100$ eV $< k\, T < 350$ eV. Others are dominated by a spectral component that
is even softer, but also include a small hard component.
Nevertheless, some sources chosen as QSSs could be intrinsically very soft
but also heavily absorbed. Any such source luminous enough to
allow a spectral fit can be reclassified as an SSS.  

It is not likely that QSSs correspond to new and unanticipated physical
systems. They probably represent extensions  of the parameter space
of sources which are already known. For example QSSs could be neutron stars
or accreting black holes in soft states, or unusually soft luminous
supernova remnants. Their effective temperatures correspond to what is
expected from the inner disks of black holes with $M\sim 100\, M_\odot.$  
They are new in that they fill the gap between ``canonical", hard X-ray sources 
and SSSs.  

\subsection{This paper}

Source detection is the subject of \S 2. 
The process we used to select VSSs is described in \S 3. 
In \S 4 we present the spectra of 
the VSSs from which we have collected the
most counts. \S 5 is devoted to optical IDs, while 
the variability of the VSSs is examined in \S 6. 
The locations 
of the sources relative to large-scale galaxy 
features and relative to other stellar populations
is discussed in \S 7.  
In \S 8 we focus on the sources with somewhat harder spectra, 
the QSSs, and in \S 9 we discuss the low-luminosity sources. 
\S 10 is a summary of our conclusions.

\section{Observations and Analyses}

The data were taken as part of two separate observing campaigns.
One was a survey studying $3$ distinct regions of the disk. The 
regions encompass the coordinates
of $9$ {\it ROSAT-}detected SSSs; we were able to place the S3 CCD,
which is especially sensitive to soft X-radiation, over regions including
$5$ of these SSSs.   
These $3$ disk fields, which span a wide range of stellar populations,
were each observed $3$ times (15 ks for each ACIS-S observation)
by \chandra\ during 2000--2001,
at intervals of 3--4 months.   

The second campaign led to superb coverage of the central regions.
The central region of M31 was observed by \chandra\ ACIS-I eight times
from 1999 to 2001, with exposure times ranging from 4 to 8.8 ks. The details of
the observations are given in Kong et al. (2002). 
The same region was also observed by \chandra\ ACIS-S for 37.7 ks
(after rejecting high background period) on 2001 October 5. We 
make use of this deep \chandra\ observation to search for SSSs near the 
nucleus. Since the soft X-ray sensitivity is best in S3, we here limit
our discussion on the nuclear region to sources detected within this    
$8'\times8'$ region.

For each 
observation, we examined the background and rejected all high-background
intervals. Only events with photon energies in the range of 0.1--7 keV
were included in our analysis. 
For the three disk fields, the three
observations in each field were merged, for all chips, to increase the signal-to-noise
ratio. To detect sources we used CIAO
task {\it WAVDETECT} (Freeman
et al. 2002). Source count rates were determined via aperture photometry
and were corrected for
effective exposure and vignetting. The radius of the aperture was varied
with the average off-axis
angle to match the 90\% encircled energy fraction. Background was extracted
from an annulus centered on each source. In
some cases, we modified the extraction
region to avoid nearby sources. It was also necessary to modify the
extraction radius for some faint sources close to more luminous
sources. Every extraction region was examined carefully in the image. 
All detected sources have signal-to-noise ratio (S/N) $> 3$, with a minimum
of 9 counts. 

For more details, see Kong et al. 2002. 
The X-ray sources found in M31's globular clusters (GCs) are 
discussed in \rd\ et al. 2002a.
A comparative study of the luminosity functions of each region has been
carried out (Kong et al. 2003). A preliminary report on the 
SSSs in M31 can be found in \rd\ et al. 2002b.

\section{Selection of Soft Sources}

\subsection{Background}

We used PIMMS to predict that each of the five disk SSSs covered by S3 
would be detected in a 15 ksec {\it Chandra} ACIS-S observation.
Our confidence 
in these predictions was bolstered
by the fact that we had completed an AO1 {\it Chandra} program
(PI: Murray) that observed  most of the local (Galactic and
Magellanic Cloud) SSSs, and that those data had been in almost perfect
agreement with the PIMMS predictions.\footnote{
The count rates are almost exactly as predicted. An analysis of these data
has not yet, however, been published. This is because the {\it ACIS-S}
low-energy 
calibration is not yet well-enough understood. At the lowest energies, 
anomalies appear in the spectra of all nearby SSSs. We do not collect
enough counts from most extragalactic SSSs for the artifacts to be apparent,
but if the calibration is completed, low-count spectra, like the ones
we present in \S 4, should
be checked.}  
None of the {\it ROSAT} sources were detected 
during the first set of {\it Chandra} observations; nor were any
 new SSSs 
with count rates comparable to
those expected for the {\it ROSAT}-discovered sources were discovered.
While there were  many very soft   
sources, many of these were foreground stars. Furthermore, 
most of the soft sources produced a small number
of counts and, typically, one or more photons had energy greater than
$1.1$ keV. 
Especially because the {\it ROSAT} PSPC had a lower-energy
cut-off than {\it Chandra}, making spectral comparisons
difficult, it was not clear which of the {\it Chandra}-discovered
sources  
should be designated as SSSs.

We therefore developed an algorithm to select SSSs from
among the characteristically low-count-rate sources in external
 galaxies (\rd\ \& Kong 2003a, b). At this point the 
algorithm, motivated by our M31 data,
 has been applied to simulated data and to {\it Chandra} data from
approximately $20$ galaxies. (See \S 1.4.) including M101, M83,
NGC 6947, M51 (\rd\ \& Kong 2003b, c), NGC 4472, M87 (Friedman et al. 2002), 
and M104 (\rd\ et al. 2003), and to {\it XMM} data from
NGC 300 (Kong \& \rd\ 2003) and from the halo of M31 
(\rd\ \& Kong 2003d). It successfully selected
very soft sources, about a dozen of which have high enough
 luminosities that  their spectra can be computed
to confirm that the spectral models are dominated by
a highly luminous soft component. 
The algorithm is described in detail in Di\, Stefano \& Kong 2003a.
Below we provide a brief sketch.
 
\subsection{Algorithm to select soft sources}
 The first step is to impose a set of strict hardness ratio conditions. 
We use
$3$ energy bins:
{\bf S:} 0.1-1.1 keV, {\bf M:} 1.1-2 keV, {\bf H:} 2-7 keV.
We consider $2$
hardness ratios, HR1 and HR2, demanding that
\begin{equation}
{{HR}1} ={{M-S}\over{M+S}} < -0.8
\end{equation}
and
\begin{equation}
HR2 ={{H-S}\over{H+S}} < -0.8.
\end{equation}
These conditions imply that
$S > 9\, M$, 
and $S > 9\, H$.
The so-called ``HR'' conditions consist of the above conditions, plus
$2$ additional criteria: $(S + \Delta S) > 9\, (M + \Delta M)$,
and $(S + \Delta S) > 9\, (H + \Delta H)$, where 
$\Delta S$, $\Delta M$, and $\Delta H$ are the one-$\sigma$ uncertainties in
$S$, $M$, and $H,$ respectively.
Sources satisfying these $4$ conditions are denoted ``SSS-HRs''.
The designation SSS-HR requires that
 the $S$ band receive
more than $13.1$   photons if no photons arrive in either the $M$
or $H$ bands,
while $S$ must receive more than $24$ photons if there is even a
single photon in either $M$ or $H.$ (See \rd\ \& Kong 2003a, b for details and 
applications to both simulated and real data.)

Sources with identical spectral characteristics to those satisfying the HR 
conditions may not be able to satisfy these conditions
if their count rates are low,
especially if one or more photons falls in the  $M$ or $H$ bands. 
We can successfully identify some such sources by relaxing the
conditions. If a source satisfies only conditions (1) and (2),
it is an SSS which we designate
``$3\, \sigma$''. 

It makes sense to relax the conditions further. Consider for example,
a source with $100$ eV, which could correspond to a nuclear-burning
WD with mass close to the Chandrasekhar limit. Such a source does emit 
some photons above $1.1$ keV.
If it lies behind a large gas column, emission in
the $S$ band can be significantly eroded, allowing the hardness
ratio, HR$_1$, to have a large value. 
The degradation of ACIS-S's soft photon sensitivity also
hardens the apparent spectrum, while sources in chips other
than S3 may also appear harder. 
That is, even if we are primarily interested in sources that 
have spectra  like those of the first batch of SSSs 
discovered in the Galaxy and Magellanic Clouds, we need to 
loosen the selection criteria and risk also selecting sources 
that are somewhat harder. 
The selection of harder sources could be 
advantageous, even for the selection of 
nuclear-burning WDs, since so far we have only $9$ candidates
for the model.   
It is unlikely that 
the spectral properties of these $9$ sources span the full gamut of
model properties. 
It is feasible, e.g., that in some WD systems, hot coronae or 
interactions with a dense interstellar medium could produce a small
hard component. 
In addition, 
if some SSSs are neutron star or black hole systems, they may occasionally
emit harder radiation or even exhibit a power law tail that carries
a small fraction of the energy.

We will therefore call a source a classical supersoft source
if it provides enough counts to allow a spectral fit
and 
has a spectrum consistent with the SSSs identified in the Galaxy
and Magellanic Clouds (Greiner 2000).  
Since we cannot fit spectra for most X-ray sources in
external galaxies, we designed the algorithm mentioned above
(\rd\ \& Kong 2003a, b).
Our algorithm consists of a sequence of selection  procedures, 
starting with those that select SSS-HR and SSS-$3 \sigma$ sources,
the sources almost certain to be similar to local SSSs.

Beyond these first two steps in our hierarchical algorithm
are a further set of seven steps. Sources selected by these
further steps are called QSSs.  The seven steps follow two conditional
branches. The first branch leads to four conditions, each of which can only
be satisfied if there is no significant detection of photons above
$2$ keV (see \rd\ \& Kong 2003a). QSSs selected along this branch are called
QSS-NOH, QSS-MNOH, QSS-SNOH, or QSS-FNOH, depending on the specific set
of criteria that selected them.  
The second branch leads to three conditions, which place strict limits
on the fraction of photons in the $H$ band relative to the $M$ and $S$ bands;
QSSs selected along this branch are called 
QSS-HR$_1$, QSS-$3\sigma_1$, or QSS-$\sigma$.

When tested on simulated 
and real data
(\rd\ \& Kong 2003a, b),
all $9$ steps selected 
sources with spectra similar to those of the
SSSs that have been discovered in the Magellanic Clouds 
and in the Milky Way (i.e., the classical SSSs), although
most such soft sources without large absorbing columns were
SSS-HRs or SSS-$3\sigma$s.  
In addition, however, each
step (including HR and $3\, \sigma$) 
also selected some harder sources, with the last $7$ steps most
likely to select sources with $100$ eV $< k\, T_{eff} < 350$ eV,
or with power-law indices between roughly 3 and 3.5.
If, therefore, we could collect enough photons 
from each extragalactic X-ray source to fit a
spectrum, some of our sources in each category
would satisfy the definition for classical SSSs, and
others would be somewhat harder, quasisoft sources.

Note that, although the selection of quasisoft sources was not 
one of our original goals, these sources are likely to be very interesting 
in their own right, as most do not seem to correspond to known
classes of X-ray sources. 
Note also that the distinction between SSSs 
and QSSs is phenomenological and 
not physical. Some nuclear-burning white dwarfs, e.g., may be QSSs,
if there is significant upscattering of photons emitted by the WD, or
if the emission emanates from a limited portion of the surface.  
Some accreting intermediate-mass black holes may be SSSs, while others
(somewhat less massive) may be QSSs. The
notion of QSSs is introduced to recognize that the VSSs
selected by our algorithm encompass a wider group
of sources than the ``classical" SSSs found locally. 
All VSSs selected by our algorithm are listed in Table 1.
We find that VSSs comprise approximately $5-15\%$ of all
the sources in each field. No GC sources are VSS, so the fraction
of non-GC X-ray sources that are VSSs can be somewhat larger. 
 
In Table 1 we list all of the sources selected by our
algorithm. The procedure that selected this source is
listed in column 10. This will allow
future studies to focus on any set of sources
deemed appropriate. 
See \rd\ \& Kong (2003a, b) for a definition of each category.
Note, however, that the
selection criteria alone cannot be used to 
distinguish between classical
supersoft sources and quasisoft sources.
The designation ``SSS" must therefore include both.

\begin{table*}
\caption{VSS source list}
{\centering
\footnotesize
\begin{tabular}{lccccccccccccc}
\hline
\hline
\multicolumn{1}{c}{Object}& R.A.& Dec. & \multicolumn{2}{c}{Soft} &
\multicolumn{2}{c}{Medium} & \multicolumn{2}{c}{Hard} & HR1$^a$ & HR2$^b$ & Count
&Category & $d$\,($''$)\\
 & (h:m:s)& $(^{\circ}:\arcmin:\arcsec)$ &Counts  &  S/N &Counts  & S/N&
Counts & S/N && & Rate ($10^{-3}$) & &\\
\hline
s2-28&00:38:18.9&+40:15:33.5&2.28&1.02&11.05&3.22&0.68&0.27&0.66&-0.54&0.35&QSS-FNOH&53\\
s2-27&00:38:23.9&+40:25:27.9&28.51&5.33&23.37&4.37&2.72&0.72&-0.1&-0.83&1.38&QSS-HR$_1$&113\\
s2-62&00:38:38.7&+40:15:11.3& 9.70&2.93&1.03 &0.43&0.00&0.00&-0.81 &-1.0&0.27 &QSS-NOH&74\\

s2-26&00:38:40.6&+40:19:57.7&45.04 &6.53 & 0.00 & 0.00 &1.23 & 0.47&-1.0&-0.95&1.17& SSS-HR&59\\
s2-46&00:38:02.9&+40:08:26.3&43.17 &5.93&6.17 &1.87&6.30& 1.47&-0.75 & -0.75 &1.40&SSS-$3\sigma$&83\\
s2-29&00:38:14.0&+40:15:22.9&36.06 & 5.90&4.32 & 1.80&0.00 & 0.00&-0.79 & -1.0 &1.02&SSS-$3\sigma$&57\\
s2-10&00:38:25.7&+40:17:39.4&8.70&2.80&0.92&0.39&1.52&0.66&-0.81&-0.70&0.28&QSS-$3\sigma_1$&33\\
s2-7&00:38:31.2&+40:17:12.0 & 31.52 & 5.57&3.09 & 1.40&0.00 & 0.00&-0.82 & -1.0 &0.87&SSS-$3\sigma$&60\\
s2-37&00:39:38.7&+40:11:00.4&190.41&11.22&36.79 &4.90&0.00 & 0.00&-0.68 &-1.0 &5.74&QSS-MNOH&120\\
\hline

s1-42&00:41:35.6&+41:06:56.8&57.75&7.36&8.10&2.37&5.45&1.74&-0.75&-0.83&1.86&SSS-$3\sigma$&70\\
s1-41&00:41:36.5&+41:00:17.6&8.44&2.60&3.88&1.67&0.74&0.30&-0.37&-0.84&0.34&QSS-NOH&50\\
s1-27&00:41:39.9&+41:04:25.7&18.23&3.71&7.92&2.49&0.00&0.00&-0.39&-1.00&0.68&QSS-SNOH&27\\
s1-69&00:41:41.9&+41:07:16.7&32.63&5.07&0.00&0.00&4.96&1.20&-1.00&-0.74&0.98&SSS-$3\sigma$&70\\
s1-18&00:41:49.2&+40:56:43.8&68.00&8.16&0.75&0.30&1.49&0.63&-0.98&-0.96&1.83&SSS-HR&55\\
s1-45&00:41:18.5&+40:52:00.0&393.10&19.66&189.17&13.55&25.36&3.95&-0.35&-0.88&15.91&QSS-$\sigma$&120\\
s1-20&00:41:43.5&+41:05:05.4&228.38&14.88&37.94&5.90&2.39&0.88&-0.72&-0.98&7.03&QSS-HR$_1$&56\\
r3-122&00:42:12.8&+41:05:58.9&45.60&6.75&0.00&0.00&0.00&0.00&-1.00&-1.00&1.19&SSS-HR&50\\
\hline
n1-31&00:45:58.1&+41:35:02.2&23.57&4.50&21.68&4.21&0.00&0.00&-0.04&-1.00&1.18&QSS-FNOH&68\\
n1-48&00:46:04.1&+41:49:42.7&25.39&4.76&2.36&1.11&0.96&0.37&-0.83&-0.93&0.75&$SSS-3\sigma$&108\\
n1-29&00:46:14.6&+41:43:18.3&5.49&2.02&2.75&1.25&0.49&0.18&-0.33&-0.84&0.22&QSS-NOH&25\\
n1-8&00:46:16.7&+41:36:56.0&61.95&7.40&14.79&3.44&3.65&1.22&-0.61&-0.89&2.10&QSS-$\sigma$&92\\
n1-46&00:46:23.7&+41:37:51.6&10.85&2.80&2.86&1.26&1.11&0.41&-0.58&-0.81&0.38&QSS-NOH&82\\
n1-66&00:47:33.3&+41:35:11.6&138.57&8.02&26.15&3.07&0.00&0.00&-0.68&-1.00&4.31&QSS-FNOH&154\\
n1-15&00:46:04.6&+41:41:23.7&20.24&4.41&8.34&2.80&0.48&0.18&-0.42&-0.95&0.76&QSS-SNOH&63\\
n1-13&00:46:05.7&+41:43:04.7&28.63&5.28&6.28&2.37&0.75&0.30&-0.64&-0.95&0.93&QSS-SNOH&28\\
n1-2&00:46:29.1&+41:43:13.9&33.65&5.74&4.16&1.80&3.63&1.54&-0.78&-0.81&1.08&SSS-$3\sigma$&61\\
n1-26&00:46:39.0&+41:39:07.5&19.23&4.27&7.66&2.48&1.64&0.57&-0.43&-0.84&0.74&QSS-HR$_1$&115\\

\hline

r2-42&00:42:36.5&+41:13:50.0&33.75&5.59&7.79&2.61&0.00&0.00&-0.62&-1.00&1.10&QSS-SNOH&45\\
r2-54&00:42:38.6&+41:15:26.3&17.55&3.46&0.51&0.18&0.09&0.04&-0.94&-0.99&0.48&SSS-$3\sigma$&24\\
r2-62&00:42:39.2&+41:14:24.4&8.91&2.32&9.88&3.02&0.66&0.26&0.05&-0.86&0.51&QSS-FNOH&5\\
r1-35&00:42:43.0&+41:16:03.9&130.94&10.99&4.58&1.77&0.00&0.00&-0.93&-1.00&3.60&SSS-HR&8\\
r2-60&00:42:43.9&+41:17:55.5&135.71&11.52&0.00&0.00&0.00&0.00&-1.00&-1.00&3.60&SSS-HR&19\\
r1-9&00:42:44.3&+41:16:07.3&675.26&25.96&235.72&15.35&47.00&6.82&-0.48&-0.87&25.47&QSS-$\sigma$&$\sim1$\\
r2-65&00:42:47.0&+41:14:12.4&20.44&4.14&0.96&0.40&0.00&0.00&-0.91&-1.00&0.56&SSS-$3\sigma$&23\\
r2-61&00:42:47.3&+41:15:07.2&95.29&9.33&0.77&0.29&1.39&0.59&-0.98&-0.97&2.59&SSS-HR&18\\
r1-25&00:42:47.8&+41:15:49.6&188.58&13.43&15.82&3.65&0.00&0.00&-0.85&-1.00&5.43&SSS-HR&16\\
r2-66&00:42:49.0&+41:19:47.1&6.64&2.30&2.56&1.12&0.62&0.24&-0.44&-0.83&0.26&QSS-FNOH&64\\
r2-56&00:42:50.4&+41:15:56.2&40.41&5.68&0.71&0.26&0.80&0.30&-0.97&-0.96&1.11&SSS-HR&26\\
r2-12&00:42:52.4&+41:15:39.7&2593.80&50.85&11.52&3.15&0.14&0.06&-0.99&-1.00&69.29&SSS-HR&26\\
r2-63&00:42:59.3&+41:16:42.8&115.79&10.64&0.00&0.00&0.00&0.00&-1.00&-1.00&3.07&SSS-HR&37\\
r3-11&00:43:14.3&+41:16:50.1&18.31&3.37&15.48&3.51&0.74&0.21&-0.08&-0.92&0.91&QSS-FNOH&23\\
r2-19&00:42:43.2&+41:13:19.2&384.01&19.55&89.68&9.44&9.29&2.91&-0.62&-0.95&12.84&QSS-$\sigma$&30\\
r3-115&00:43:06.9&+41:18:09.0&50.49&6.70&0.00&0.00&0.00&0.00&-1.00&-1.00&1.34&SSS-HR&37\\

\hline
\end{tabular}
\par
\medskip
\begin{minipage}{0.8\linewidth}
\footnotesize

$^a$ $HR1=(M-S)/(M+S)$\\
$^b$ $HR2=(H-S)/(H+S)$\\

{\bf Column 1:} source name. 
{\bf Columns 2 \& 3:} right ascension and declination. 
{\bf Columns 4, 5, \& 6:} the corrected numbers 
of soft, medium and hard X-ray counts detected for each source. Source count rates were determined via aperture photometry
and were corrected for effective exposure and vignetting.
{\bf Columns 7 \& 8:} hardness ratios, see \S 3.2.
{\bf Column 9:} the corrected number of counts per second.
{\bf Column 10:} selection criterion by which the source was selected as an SSS.
(See \rd\ \& Kong 2003a, b.)
{\bf Column 11:} angular distance to the nearest X-ray source.\\
Note that the naming conventions are explained in Williams et al.\, 2004. 
\end{minipage}
\par
}
\end{table*}

\section{Spectra}
We examine the energy spectra of bright SSSs. Only those in the
central region provide enough counts for a reasonable spectral fit. 
Only photons with energies between $0.3$ keV and $7$ keV
were used. Figure 2
shows four representative spectra. It is worth noting
that the degradation of soft energy ($< 1$ keV) sensitivity might affect
the spectral fits significantly and
we therefore correct the response matrices to take this
effect into account. r2-12 is the brightest SSS in M31 and is seen in every
\chandra\ observation; it was also detected by previous \rosat\
observations (Supper et al. 1997, 2001; Primini et al. 1997). A single
blackbody model cannot provide a good fit [$\chi^2_{\nu}=2.11$ for 33 
degrees of freedom (dof)]; there is an excess above
$\sim 1$ keV. From archival \xmm\ observations, r2-12 shows a high energy
tail up to $\sim 5$ keV (Kong et al., in preparation). We therefore add an
additional powerlaw component (fixed at $\alpha=2$) to improve the fit 
($\chi^2_{\nu}=1.73$ for 32 dof with $>99\%$ significance for an 
additional power-law component). 
The powerlaw component contributes only about 0.3\% of the total flux. 
The fit is not sensitive to the choice of photon index; 
we fixed it with different values from 1 to 3 and it did not 
significantly change the result. 
For r1-9, the spectrum also requires a two-component model (blackbody +
powerlaw) to achieve a good fit ($\chi^2_{\nu}=0.96$ for 31 dof). 
It is worth noting that r1-9 is only about $1''$ from the galaxy center 
(Kong et al. 2002); 
in this crowded region, it is difficult to eliminate the possibility of 
contamination from nearby sources. However, we do not know if this is the 
source of the residuals around 0.6--0.8 keV.
r1-25 is fit by a
blackbody model with $kT=122$ eV ($\chi^2_{\nu}=1.03$ for 6 dof), while 
r2-60
is well fit by a model with $kT=25$ eV ($\chi^2_{\nu}=0.63$ for 5 dof).

\section{Source IDs}

We attempted to identify the VSSs with sources observed
at other wavelengths to ($1$) search for possible counterparts,
and ($2$) distinguish between M31 VSSs and sources
that may be associated with foreground or background objects.

{\it Chandra's} good angular resolution,
 in combination
with the spectral sensitivity of ACIS-S can be crucial for this task.
Time variability can also be helpful.  
Consider two VSS counterparts suggested by previous or
ongoing work. One is a nova, associated with a {\it ROSAT}
SSS (Nedialkov et al. 2002). 
Because the spatial resolution of {\it ROSAT} makes the 
correspondence difficult to establish, the timing of the
nova and the apparently coordinated optical and X-ray decline
play a crucial role in supporting the identification.
More recently (Williams et al.\, 2003), an association 
has been suggested between
a {\it ROSAT}-observed recurrent transient, (White et al. 1995)
an X-ray source observed by {\it Chandra-HRC}, 
and an SNR. The time variability of
the X-ray source would seem to be incompatible with the SNR 
interpretation. Clearly a verification that the {\it HRC-}observed source 
is a VSS would be important.

\subsection{Search for Matches}

We have searched for correlations between the VSSs and all cataloged
objects, we have examined images from {\it The Local Group Survey} (LGS;
Massey et al.\, 2001) 
the {\it Digital Sky Survey} (DSS), and from {\it HST}.
There are a number of {\it HST} images containing SSS positions
near the center of M31, but few images of the disk.
Fortunately, the LGS and the DSS complement the {\it HST} observations,
with the combination of these $2$ optical surveys
providing good coverage of most of the disk.  

To estimate the magnitude of VSS optical counterparts, we scale
the
optical magnitude of SSSs in
the LMC (Greiner 2000) to the distance
and reddening of M31. Typical $B$ magnitudes of the LMC's SSSs 
are between
$16$ and $21$ at a distance of 50 kpc. 
Using a distance to M31 of 780 kpc, we
expect $B=22-27$ if the SSSs are like those in the LMC. 

\subsubsection{{\it HST}}

Table 3 lists the VSSs for which there is archival HST data. 
Because of the irregular WFPC2 footprint, not all
such images would necessarily cover the coordinates of the object,
so we retrieved the images from the archive, created
cosmic-ray rejected `stacked' images, and checked
the object positions in each image. We found that images
taken with the UV F160BW filter were essentially blank
everywhere, with no indication of stars or even the
galaxy nucleus, so we did not consider them further.
The HST datasets in which the supersoft sources appear are shown in Table 3.

To identify possible optical counterparts to the supersoft
sources, we used DS9 to mark the object positions on the
images, and then visually inspected them. (The regions we considered
around each source were circles enclosing $90\%$ of the 
energy.) 
Since all of the positions were close to the center of M31,
the error circles contain many faint stars, any (or none) of which
could be the true optical counterpart of the supersoft source.
Almost none of the error circles contained optical sources
which were significantly brighter than the typical stellar background.
To quantify the detection limits, we measured the background
$\sigma$ at the location of each object in each WFPC2 image
and computed the magnitude (in the WFPC2 bandpass ``Vegamag'' system)
corresponding to an object with a $20\sigma$ flux. The resulting
limits appear in Table 3.

Two supersoft sources were detected in HST images. r1-35
appears as an optical transient in the HST datasets
u2lg020 and u2lh010; 
these data were obtained in
June 1995. No optical transient appears in other images
which cover the same position, including u2c7010 (September 1994),
u2e2010 (October 1994), and u5lt010 (February 2000).
The optical magnitudes of the transient are given in Table 3.
(See also Kaaret 2002.)

r2-56 was identified with planetary nebula 462 in the catalog of
Ciardullo et al. (1998), but, since it is resolved
at X-ray and radio wavelengths, it is more likely 
to be a supernova remnant (SNR;
Kong et al. 2003) Although
its coordinates are located within many of the HST images,
it is only visually apparent in a narrow-band H$\alpha$ image
(see Figure~3). The H$\alpha$ magnitude is given in Table 3.

\begin{table*}
\caption{VSS Variability and Optical IDs}
{\centering
\footnotesize
\begin{tabular}{lccccc}
\hline
\hline

Source & Optical ID  & S-Value & {\it Chandra} & {\it XMM} & Note\\
\hline

s2-28 & $V > 21.6$ & 1.09 & & \nodata &\\
s2-27 & $V > 21.8$ & \nodata & & \nodata &\\
s2-62 & \nodata& 2.53 &t & \nodata &\\
s2-26 & $V > 21.6$ & 3.17 & v& \nodata&{\it ROSAT}\\
s2-46 & $V<15$ & 3.13 & v&\nodata & saturated star\\
s2-29 & $V<15$ & 0.12 & &\nodata & saturated star\\
s2-10 & $B=21.1$;$V=19.4$ & 1.98 & & \nodata & possible symbiotic\\
s2-7 & $B=16.3$;$V=15.6$ & 1.69 & &\nodata & \\
s2-37 & $V<15$ & \nodata & & \nodata&saturated star\\
\hline

s1-42 &  \nodata & 2.71 & & on &SNR \\
s1-41 &  \nodata & 0.75 &  & on & Globular cluster? \\
s1-27 &  \nodata & 3.24 & v,t & off & \\
s1-69 &  \nodata & 2.01 & t& off &\\
s1-18 &  \nodata& 7.72 & v,t & off & \\
s1-45 & $B=13.4$;$R=12.2$ &  0.51 & & on & USNO star\\
s1-20 & $B=12.2$;$R=11.2$ & 2.36 & & on & USNO star\\
r3-122 & $B=18.7$;$R=18.4$ &  \nodata & & off & USNO star\\

\hline
n1-31 & $V > 21.8$ & 0.32 & & \nodata &\\
n1-48 & $V > 21.8$& 1.06 & & on & \\
n1-29 &  \nodata & 0.27 & & \nodata &\\
n1-8 &  \nodata & 1.23 & & \nodata &\\
n1-46 & $V > 21.8$ & 1.18& & \nodata & \\
n1-66 & \nodata & \nodata & & \nodata &\\
n1-15 & $B=19.1$;$V=17.6$  & 1.13&& \nodata &\\
n1-13 & $B=20.2$;$V=18.7$  &1.62& & \nodata &\\
n1-2 & $B=19.9$;$V=18.4$  &1.57& & \nodata &\\
n1-26 & $B=19.0$;$V=17.6$ &3.01& v,t & \nodata &\\

\hline
r2-42 & \nodata& 2.32 & & on$^b$ &\\
r2-54 &  {\it HST} & 1.53 & & off &\\
r2-62 &  \nodata& &  t & off\\
r1-35& {\it HST} & &t & off &S And$^{3}$\\
r2-60 & $V > 19.5$,{\it HST}& &t & on$^a$ &\\
r1-9 &  \nodata & 7.16 & v,t & \nodata&unresolved with {\it ROSAT} and {\it XMM}\\
r2-65 &  \nodata & & t & off\\
r2-61 &   {\it HST}& &  t & off\\
r1-25& {\it HST} & 2.83 & & off & \\
r2-66 & $V > 19.5$& & t& off\\
r2-56 & {\it HST}&  1.71 & & off$^c$ &SNR$^2$\\
r2-12 & {\it HST} & 5.92 & v & on$^b$ & {\it ROSAT}\\
r2-63 & $V > 19.5$& & t& off\\
r3-11 &$V > 20.1$& 2.16 & & off &\\
r2-19 & $I=18.3$$^1$ &  2.56 & & on$^b$ & \\
r3-115 & $B=23.3$;$V=22.0$ &  & t& off&possible symbiotic\\

\hline

\end{tabular}
\par
\medskip
\begin{minipage}{0.8\linewidth}
\footnotesize
Note. --- Except s1-20, s1-45, r3-122, and r2-19, photometry of stars is derived from the LGS. For the {\it HST} observations, please refer to Table 3 for the detection limit.\\ 
References. ---$^1$ Haiman et al. 1994; $^2$ Kong et al. 2003; $^3$ Kaaret 2002\\
$^a$ On during the 2nd two (of four) {\it XMM} observations\\
$^b$ On during each of four {\it XMM} observations\\
$^c$ Near the nucleus and an extended X-ray object. It is highly contaminated by the diffuse emission.\\

Dots indicate that no relevant information was available.
{\bf Column 1:}  source name. 
{\bf Column 2:} Optical ID; magnitudes of the USNO stars are 
taken from the USNO catalog, I magnitude of r2-19 
is from Haiman et al.\, 1994; all others are from the LGS; 
$V$-band detection limits from the LGS;   
entries marked {\it HST} should be checked in Table 3. 
{\bf Column 3:} $S$ is the variability factor defined in the text; 
$S>3$ has been taken to denote variability. 
{\bf Column 4:} Variability as determined within the {\it Chandra} 
data set; v denotes variable, t indicates that the source 
was below the detectability limit at least once. 
{\bf Column 5:} ``on'' and ``off'' indicate whether the source was 
detected or not by a visual inspection of the {\it XMM} data. 
{\bf Column 6:} Notes;   
{\it ROSAT}=detected with {\it ROSAT}, ``unresolved'' means could
not be resolved from the nuclear source.\\  

\end{minipage}
\par
}
\end{table*}

\begin{figure*}
\epsfig{file=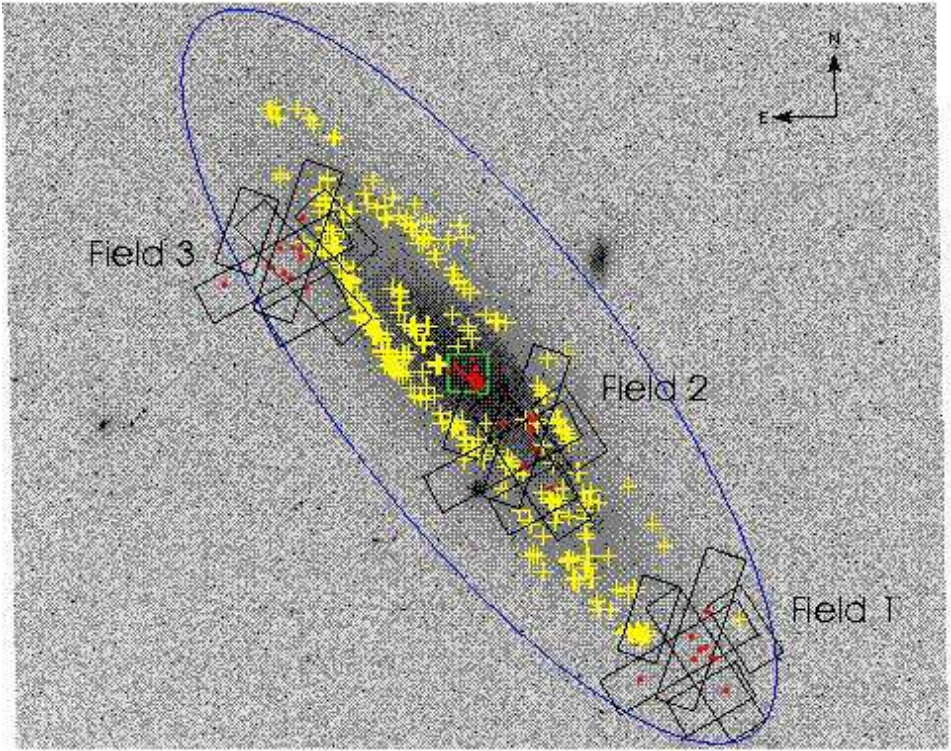,width=5in}
\caption{Detected VSSs (red dots) overlaid on an optical Digitized Sky
Survey image of M31. The fields of view of the three \chandra\ ACIS-S 
observations in
the disk (black boxes) and the central region (green box) are also shown.
Also shown in the figures are the optical position of SNRs (yellow plus
signs). The ellipse shows the $D_{25}$ isophote.}
\end{figure*}

\begin{figure*}
\centerline{
\rotatebox{-90}{\psfig{file=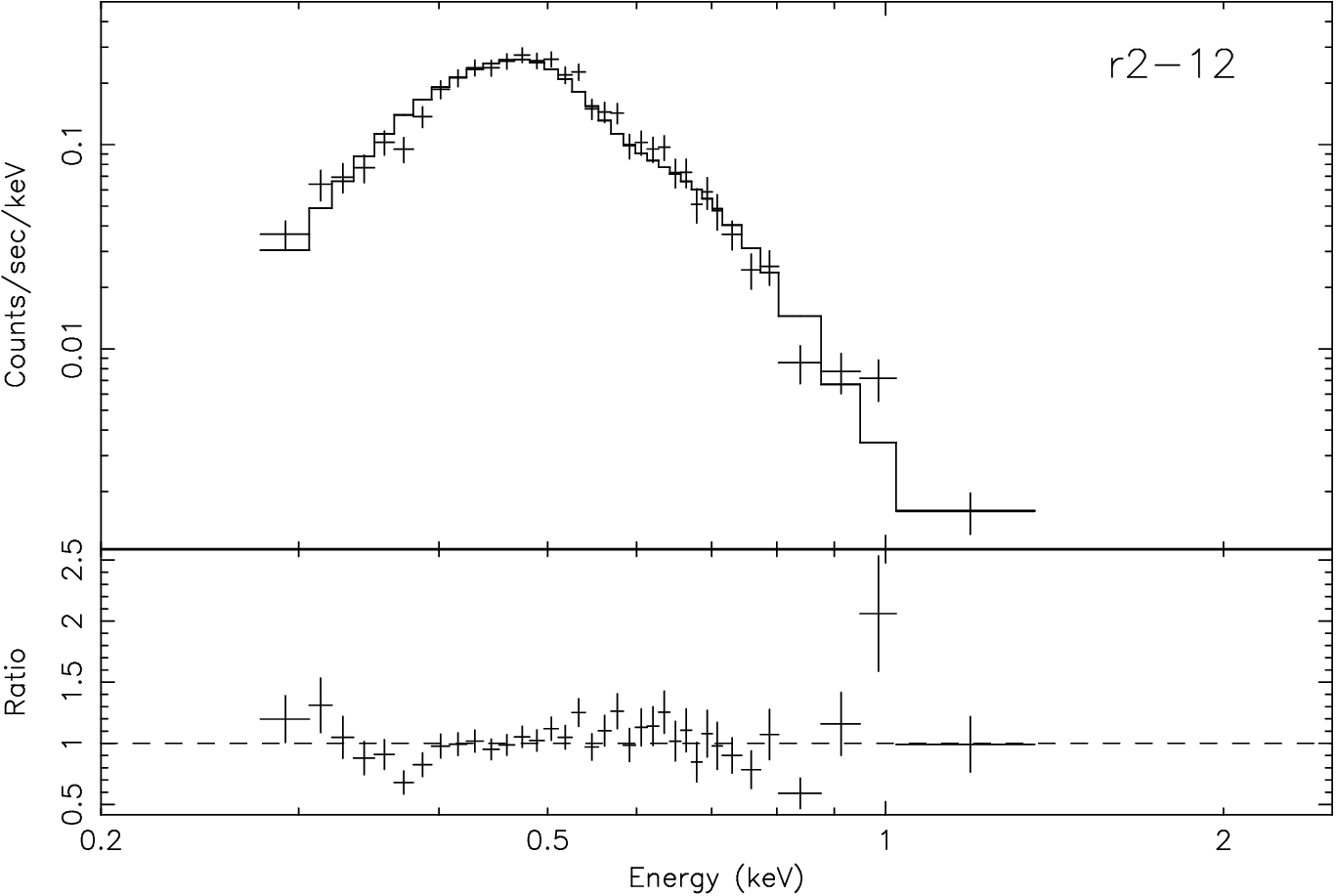,height=3.1in}}
\rotatebox{-90}{\psfig{file=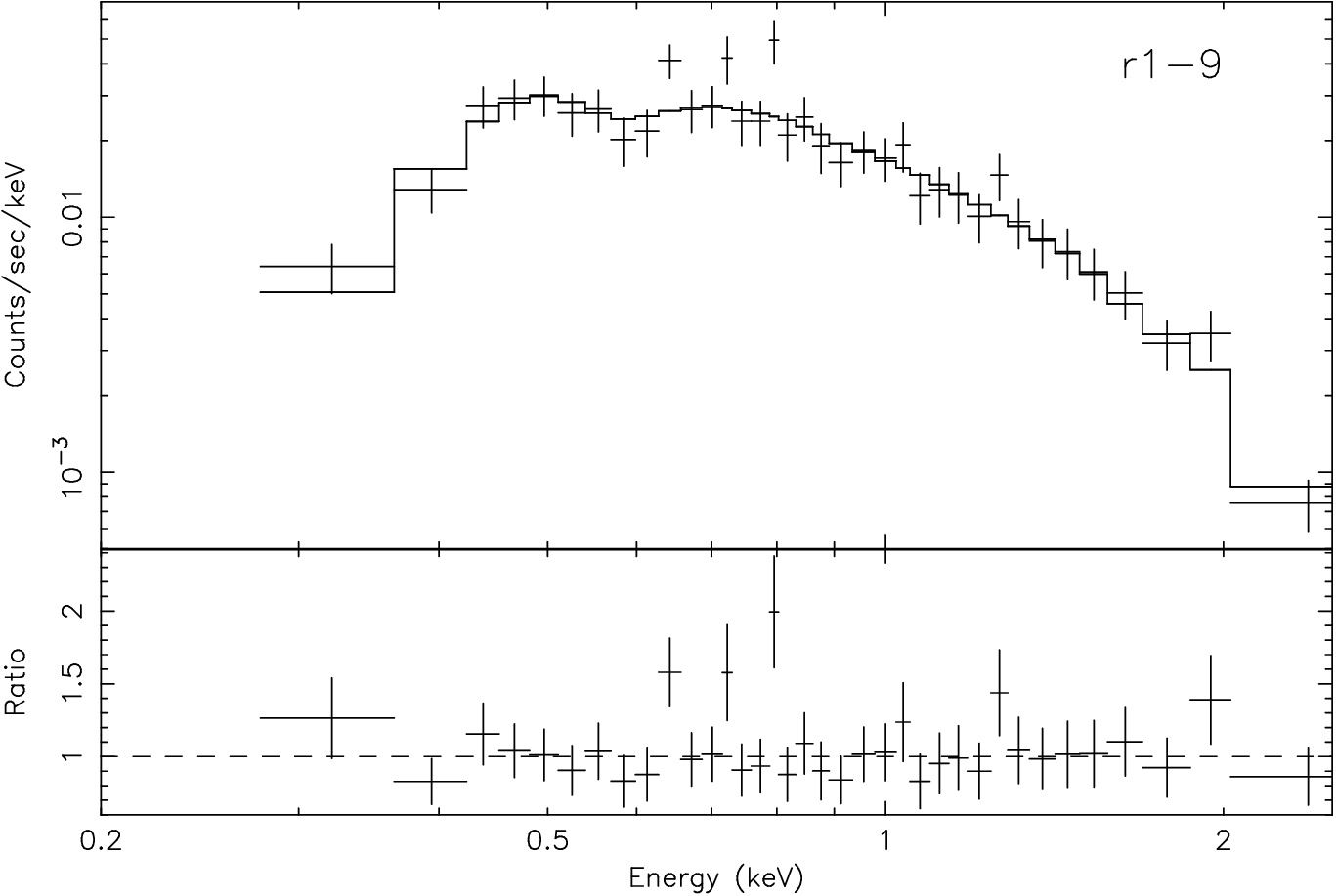,height=3.1in}}
}
\centerline{
\rotatebox{-90}{\psfig{file=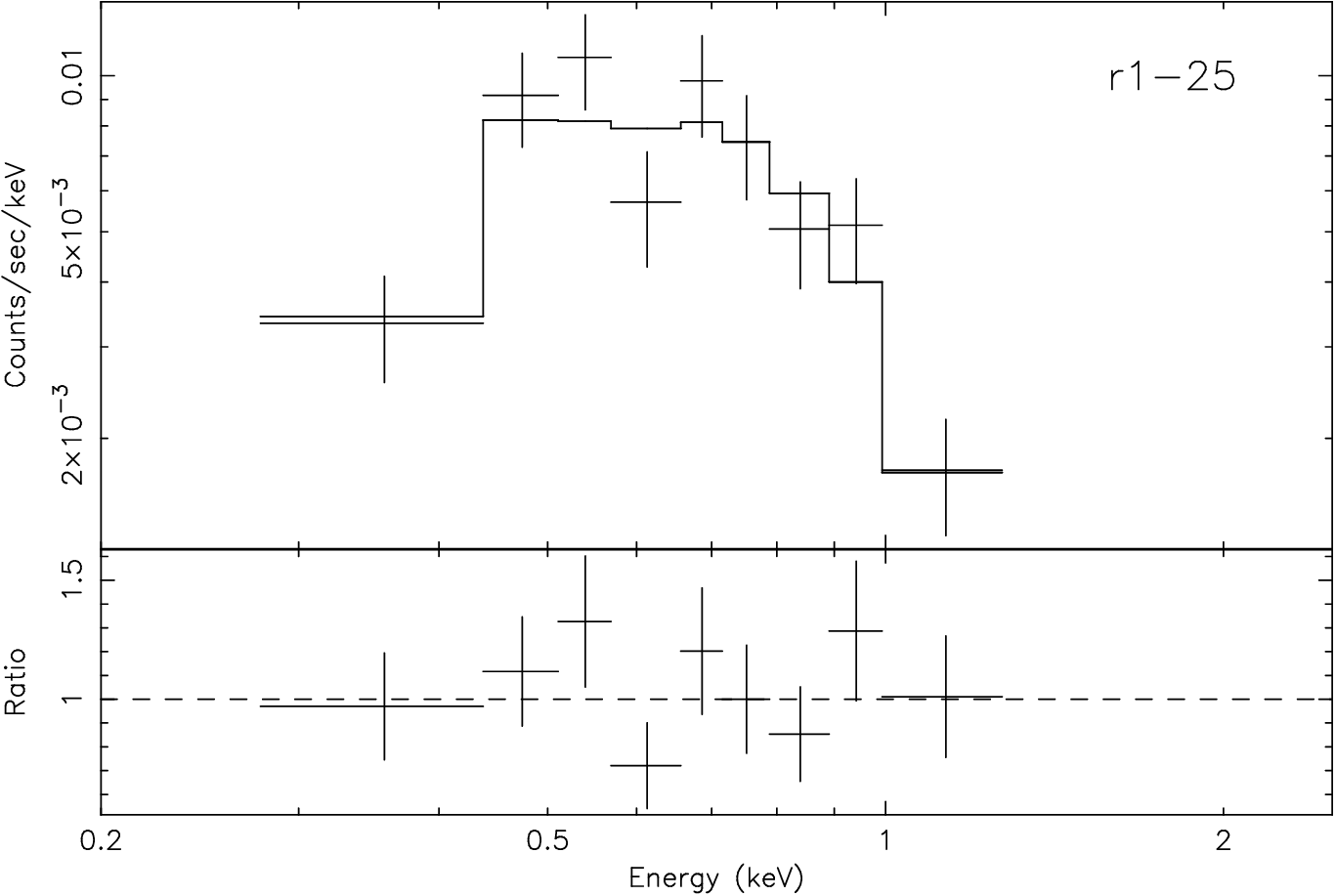,height=3.1in}}
\rotatebox{-90}{\psfig{file=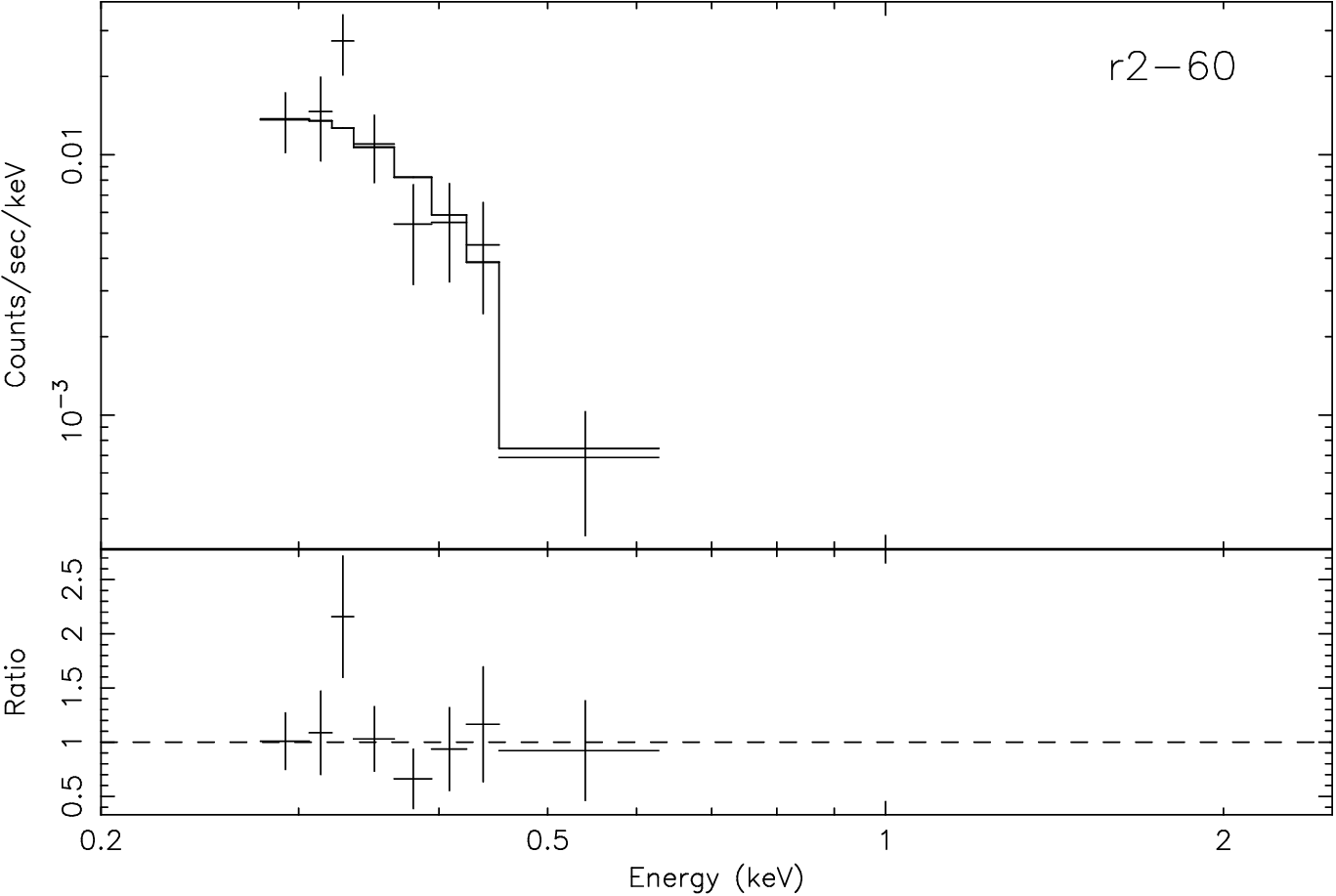,height=3.1in}}
}
\caption{Spectral fit to {\bf r2-12} (blackbody + power-law model with
$N_H=2.4\times10^{21}$ cm$^{-2}$, $kT=56$ eV and $\alpha=2$ (fixed);
$L_{0.3-7}=4.9\times10^{38}$\lum\ and $\chi^2_{\nu}=1.73$ for 32 dof), 
{\bf 
r1-9} (blackbody + power-law model with
$N_H=2\times10^{21}$ cm$^{-2}$, $kT=89$ eV and $\alpha=3.3$;
$L_{0.3-7}=3.3\times10^{37}$\lum\ and $\chi^2_{\nu}=0.96$ for 31 dof), 
{\bf 
r1-25} (blackbody
model with $N_H=1.1\times10^{21}$ cm$^{-2}$ and $kT=122$ eV;
$L_{0.3-7}=3.6\times10^{36}$\lum\ and $\chi^2_{\nu}=1.03$ for 6 dof), and 
{\bf r2-60} (blackbody mode with
$N_H=1.1\times10^{21}$ cm$^{-2}$ and $kT=25$ eV;
$L_{0.3-7}=1.5\times10^{37}$\lum\ and $\chi^2_{\nu}=0.63$ for 5 dof).}
\end{figure*}

\begin{table*}
\caption{HST data for M31 VSSs}
{\centering
\footnotesize
\begin{tabular}{lllllllll}
\hline
\hline

Source & Dataset names & F300W& F336W & F547M & F555W & F656N & F841W & F1042M\\
\hline

r1-25 & u2c7010, u2e2010, u2lg020, u2lh020 &  22.2 & 20.3 & 22.5 & 22.5 & 21.0
&21.0 &19.7\\
r1-35 & u2c7010, u2e2010, u2e2020, u5lt010 &  \nodata & 21.5 & 23.3 & 23.4 &
21.9 & 21.9& 20.0\\
      & u2lg020, u2lh020 (detection) & 17.5 & 17.5 & \nodata &18.5 &\nodata &18.0 &\nodata\\
r2-12 & u2c7010, u2lh010 &\nodata & 20.3 & 22.8 & \nodata & 21.8 & \nodata
&\nodata\\
r2-54 & u2e2020, u2kj010, u5lt010 &\nodata & 21.5 & \nodata & 22.9 & \nodata &
21.2 & \nodata\\
r2-56 & u2c7010, u2e2010, u2lg020, u2lh020 & 22.2 & 20.3& 22.7 & 23.1 &
16.9$^a$ &21.4 & 19.9\\
r2-61 & u2lg020, u2lh010 &22.2 & 20.3 & \nodata & 22.7 & \nodata & 21.2 &
\nodata\\
r2-60 & u2e2020 &\nodata &\nodata &\nodata & 23.1 &\nodata & 21.5 & 20.0\\
\hline
\end{tabular}
\par

\medskip
\begin{minipage}{0.8\linewidth}
\footnotesize

$^a$ Object is detected in this filter\\

\end{minipage}
\par
}

\end{table*}


\subsubsection{Optical Surveys}

We examined {\it The Local Group Survey} (LGS) and Digital Sky Survey (DSS)  images containing the coordinates of each SSS.
{\it The Local Group Survey} project 
uses the 4-m
telescopes of NOAO for an optical survey of all the Local Group galaxies
currently forming stars, in $UBVRI$ and the narrow-band nebular filters
Halpha, [OIII], and [SII].
These data have spatial resolution of 1 arcsec  
and go to a S/N of 10 at $U=24.5$, $B=V=R=I=25.0$,
with good astrometric ($0.3''$) accuracy.

For the DSS images, we examined the images for each source visually 
and then extracted the magnitude from the USNO catalog (Monet 1998).

\subsubsection{M31 Catalog}

Table 2 summarizes the results of matching our new \chandra\ source
catalog with existing catalogs of M31 X-ray sources (Supper et al.
1997, 2001). Due to the poorer spatial resolution of the \rosat\ PSPC, we used
a relatively large searching radius ($15''$) to cross-correlate with the
\chandra\ catalog. 

\subsection{Matches}
 
\subsubsection{Supernova Remnants}

Supernova remnants (SNRs) can have very soft X-ray spectra. 
To identify those VSSs which are SNRs, we first
checked for matches
between the VSSs and SNRs that have been previously 
identified through optical and radio
surveys. We found $2$ matches.  

Optical, radio and X-ray observations (Kong et al.\, 2003) 
have resolved r2-56, solidly establishing that it is an SNR.
Yet, this source is one of the
softest XRSs in our sample, with no emission above $1.1$ keV.
This clearly indicates that we cannot eliminate SNRs from any
sample of spectrally selected soft sources
simply by tightening the eligibility criteria to include 
only the softest sources.
The Field 2 source s1-42 is also identified as an SNR through optical observations
(Magnier et al. 1997).

Are these $2$ SNRs the only ones among our list?
Certainly they are the only ones identified through matches
at other wavelengths. But, because the relative signature
in X-ray, optical, and radio can vary, depending on the
environment of the supernova progenitor, it is possible that
some X-ray active SNRs could be missed at other
wavelengths. It is therefore important to have a
second discriminant.
Fortunately, time variability provides a second test.
Because the X-ray emission from an SNR presumably emanates from
an extended region, it should not vary significantly
(e.g., by a factor of $2$) over periods much smaller than years.
Neither r2-56
nor s1-42 satisfied our criteria for variability.

\subsubsection{Symbiotics}

In symbiotics, wind from a giant star carries mass to a
WD companion at rates high enough that nuclear burning can occur.
Of the $22$ SSSs found in the Galaxy and Magellanic Clouds,
$3$ are symbiotics. 
Morgan (1996) gives the range for symbiotics in the LMC
as V: 14.7-16.5, with B-V: 0.3-1.7.
The V magnitude roughly translates to V: 20.1-21.9 at M31 distances.
 
Stars were identified and measured in the LGS data using DAOPHOT
II/ALLSTAR (Stetson, Davis, \& Crabtree 1990) with a detection threshold of
5$\sigma$.  This objective technique detects a star within the 1" error
circle of r3-115, with V$\approx$22, B-V$\approx$1.3.
 Using HST
images, we find that a
star is detected at the  $4.6 \sigma$ level,
 at RA = 0:43:06.8669, Dec =
41:18:08.758, with $U=23.57\pm 0.24$.  The field is crowded enough that there
is about a $25\%$ chance that this is a chance match.
Although there are uncertainties about the identification in each
data set, the combination is promising.

In Field 1, which is the least crowded M31 field we have observed,
s2-10 is identified with a star in the LGS, having V=19.4,
B-V=1.7. Although this system is half a magnitude
brighter than
the range indicated by  
Morgan (1996), that range was based on the study of only $10$ systems.
s2-10 should therefore be considered as a candidate
symbiotic. 

\subsection{Foreground Stars and Other Contaminants}

Table 2 illustrates that foreground stars are significant contaminants. 
We have assumed that any bright star (generally $B < 19$) 
found at the position of
an VSS is a foreground star. Because s2-10 and r3-115
have colors and  magnitudes that place them in the range of M31 symbiotics,
we consider that they may belong to M31.
The magnitudes of some other stars in Table 2 may also
be consistent with M31 membership.
While these may therefore warrant further investigation,
for now we consider that they are likely foreground stars 

After the elimination of foreground stars,
the number of VSSs remaining in our disk fields is
small ($5-6$ each in Fields 1, 2, and 3). 
It is important to 
estimate the likely contribution of any contaminants,
such as foreground magnetic CVs, or distant luminous
background sources, that may contribute to the non-stellar
sources not already identified.

Some background objects, including some distant AGN,
may be identified at other wavelengths.
In other cases, an identification may be ambiguous.
Consider, e.g., s1-41. This is one of the dimmest
X-ray sources in our sample ($\sim 13$ counts). It is a QSS.  
This source is identified with a globular cluster, Bo 251. 
Although we knew of this identification when we published
work on X-ray sources in M31 GCs (\rd\ et al.\, 2002a), we did not
include this association in the GC list, because it is not
clear that the optical object is indeed a GC. It may be
a more distant galaxy or cluster of galaxies, or it may
be something else. 
This is the only such object we are aware of in the disk fields.

Magnetic CVs or other soft (but optically dim) XRSs located
in the Milky Way's disk or halo can be more difficult
to eliminate, especially if the donor is a low-mass star and
the disk is small or non-existent. 
 
To estimate the level of contamination from 
all non-M31 sources which have not been identified at
other wavelengths,
we have studied data analyzed by
the {\it ChaMP} collaboration. We searched their archives
for publicly available analyzed ACIS-S data on fields
that (1) have been observed for at least $10$ ksec, and (2) that
do not contain clusters (GCs or galaxy clusters), gravitational
lenses, or nearby galaxies.   
The energy bins used by {\it ChaMP} (http://hea-www.harvard.edu/CHAMP/)
are slightly different
from the ones we have used ($S_{ch}= 0.3-0.9$ keV; 
$M_{ch}= 0.9-2.5$ keV;
$H_{ch}=2.5-8$ keV). We nevertheless used them   
exactly in the manner described in \S 2 to select VSSs.
In each of the $4$ fields we studied,
we found $1-3$ sources per field that satisfy the VSS criteria.
These numbers are consistent with the
expected numbers of foreground stars. 
This indicates that VSS interlopers not visible at
other wavelengths
are rare, a result consistent with previous surveys (see, e.g.,
Becker 1997, PhD thesis).  
 
\begin{figure*} 
\psfig{file=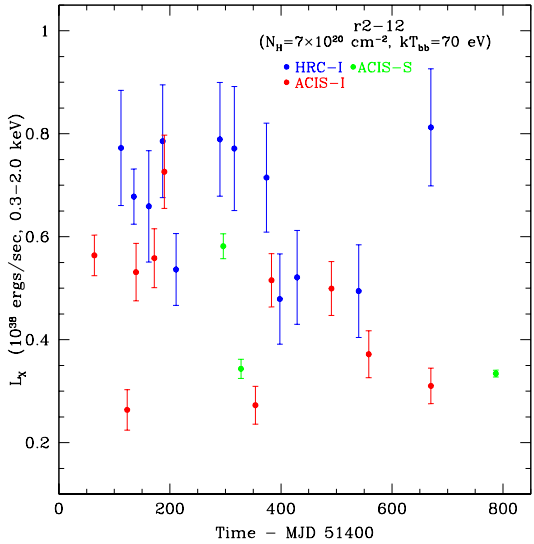,width=3in}
\caption{
Long-term light curve of the brightest VSS, the SSS-HR r2-12, spanning from  
1999 November to 2001 October. A blackbody spectral model ($kT=70$ eV 
and $N_H=7\times10^{20}$ cm$^{-2}$) is assumed to derive the 0.5--7 keV 
luminosity.
}
\end{figure*} 

\begin{figure*}
\psfig{file=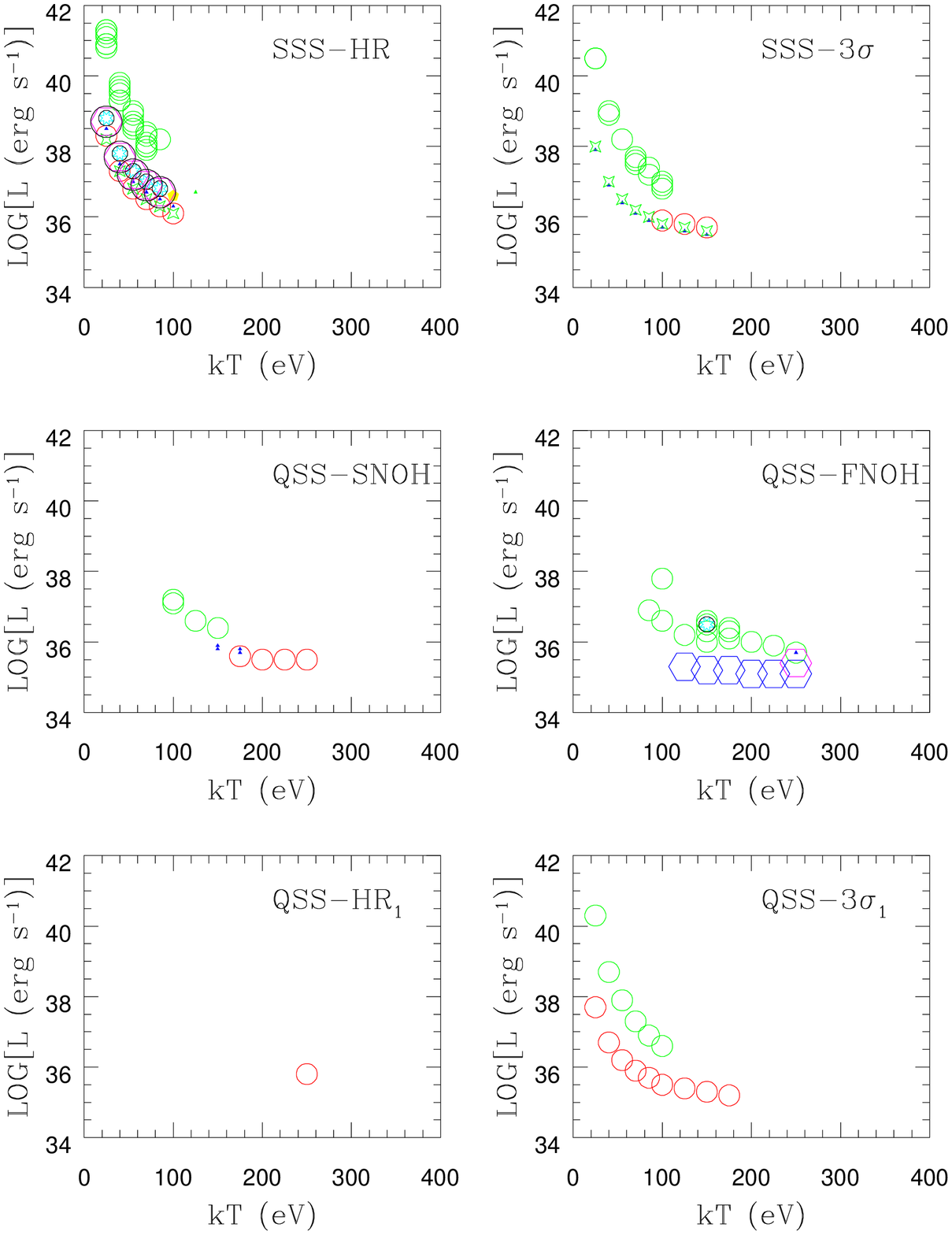,width=6in}
\vspace{1.1 true in}
\caption{
$Log(L_x)$ versus 
$k\, T$ for M31's VSSs. See text.
Each panel is labeled according to the classification
of the sources considered in that panel.
The lower curve corresponds to blackbody models with
$N_H=1.6 \times 10^{21}$ cm$^{-2}$; each source is shown with
a unique symbol.
The upper curve corresponds to blackbody models with
$N_H=6.4 \times 10^{21}$ cm$^{-2}$; one symbol is used for all sources.
}
\end{figure*}

\section{Variability} 

To quantify the level of X-ray variability, we
have taken three steps: we searched for the {\it ROSAT}-observed
SSSs in the {\it Chandra} data set; we searched for the 
{\it Chandra}-observed VSSs in the {\it XMM} data set; 
we compared the sets of VSSs observed by {\it Chandra} during
different observations of the same field. 
 
\subsection{{\it ROSAT} sources in {\it Chandra} fields}

M31 was observed by {\it ROSAT} in 1991 and 1992.\footnote{
All of the published sources were observed in 1991; any variations
between 1991 and 1992 have not been discussed in the literature.}  
{\it Chandra} observations occurred $9-10$
years afterward, so the combination provides information
about variability on the time scale of roughly one decade.

We checked the position of every known SSS observed by {\it ROSAT},
to determine if it was in a {\it Chandra}-observed field and, if so, whether
it was detected. Eight of the sources should have been detected (yielding 
20-150 counts) by {\it Chandra} during the $\sim 15$ ksec observations that we
conducted. (See Greiner et al. 2004 for more detailed 
consideration of this point.) 
\footnote{
The single {\it ROSAT} SSS that would not have been 
detected is RX J0045.4; the {\it ROSAT} count rate 
was $<10^{-5}$, and the source was far 
off-axis in the {\it Chandra} observation.} 

\smallskip

\noindent{\sl Field 1:\, } Three {\it ROSAT} sources are in Field 1.
The position of one of the sources (RX J0037.4) was observed only once, and the 
source was not detected. Two additional sources were located in S3;
each was observed $3$ times. One of these $2$ sources (RX J0038.5) was never
detected by {\it Chandra}; the second (RX J0038.6) was detected in March 2001,
but not during the other $2$ observations. 

\smallskip

\noindent{\sl Field 2:\, } {\it ROSAT} identified $1$ source in Field 2;
RX J0041.8 was located in S3, and was not detected in any of the $3$ 
{\it Chandra} observations.

\smallskip

\noindent{\sl Field 3:\, } Five {\it ROSAT} sources are in Field 3.
Of these $5$, RX J0047.6, would not
have been detectable in a $15$ ksec {\it Chandra} observation,
unless the count rate was higher than it had been during the 
{\it ROSAT} detection. We might have expected 
two other sources to be detected as very weak sources,  
(a few counts per {\it Chandra} observation), because the
{\it ROSAT} count rates were small, and the sources (RX J0045.4 and RX J0046.1) were not in S3. Two other sources (RX J0046.2+4144 and RX J0046.2+4138) would have been clearly detected if their flux had been comparable to that measured by 
{\it ROSAT}.
None of the  {\it ROSAT} sources were detected by {\it Chandra}.

\noindent{\sl Nuclear region:\, } 
{\it ROSAT} detected one SSS in the nuclear region, but the uncertainty
in its position is too large to resolve it from  $2$
VSSs detected by {\it Chandra.}

\subsection{{\it Chandra} sources in {\it XMM} fields}

M31 has been observed by {\it XMM} several times since 2000.  In particular,
the central $\sim 15'$ region was observed $4$ times (2000 June for $30$ ks, 2000
December for $12$ ks, 2001 June for $39$ ks, and 2002 January for $64$ ks). 
Four regions located
along the major axis of the disk were each
observed once in 2002 January for 60 ks (for instance, see Trudolyubov
et al. 2002). For each field, we removed high background
periods and combined all three {\it XMM} detectors (MOS1 + MOS2 + pn) into
a composite image. We then visually inspected  the
composite {\it XMM} images in each region containing a {\it Chandra} VSS.
The results are summarized in Table 2. 
Fifteen sources that had been detected by {\it Chandra} were not 
detected by {\it XMM}.

\subsection{Variability Detected by {\it Chandra}}

For all sources observed more than once by {\it Chandra} we computed a variability 
parameter following Primini et al. (1993):
\begin{equation}
S(F_{\max}-F_{\min}) =
{{\Big\arrowvert{F_{max}-F_{min}}\Big\arrowvert}\over
{\sqrt{\sigma^2_{F_{max}}+\sigma^2_{F_{min}}}}}
\end{equation}
where $F_{min}$ and $F_{max}$ are the minimum and maximum X-ray fluxes 
observed and $\sigma_{min}$ and $\sigma _{max}$ are 
the corresponding errors. Sources with $S>3$ are marked as variables, v,
in Table 2. Note that if a source provides fewer than $\sim 10$ counts 
it cannot satisfy the variability criterion, even if it is not detected 
in the other observations. Some sources were detected in one or more 
{\it Chandra} observations, but provided no counts above what was
expected from background in at least one observation. These sources
are marked with a ``t'' in Table 2.

\subsection{Summary} 

All $8$ of the {\it ROSAT}-discovered SSSs in our disk fields 
that could have been detected by {\it Chandra} are transients.

Thirteen of 16 VSSs in the central field are transient. 
One of the remaining sources (r2-12) is highly variable (see figure 3). 
In Field 2, the disk field that is closest to the nucleus, $3$ of $5$ 
non-stellar VSSs are transient. In Field 1, located far out along the 
major axis, $1$ of $5$ non-stellar VSSs is transient and one is 
variable. None of the $5$ non-stellar VSSs in Field 3 has been 
found to be transient or variable. The result could have been 
different had {\it XMM} observed a longer portion of Field 3. 
The small S values in this field may, however, suggest a lower 
level of variability than in the other fields.

\section{Location}

The location of VSSs, and the stellar populations within which we find
them,  can provide important clues as to the natures of the sources.
Old sources are likely to be scattered through the galaxy,
wandering through its disk, bulge and halo.
The standard models of SSSs would tend to predict that SSSs should be
old, or at least intermediate-age systems. This is because the 
time at which mass starts being transferred to a white dwarf at high rates
($\sim 10^{-7} M_\odot$ yr$^{-1}$), 
is generally governed by the evolution time of the donor. 
For both close binary SSSs and for symbiotics, donor masses are typically
$\sim 1.5-2\, M_\odot$ (Rappaport et al. 1994, Kenyon 1986); in the former case, the donor is slightly
evolved and in the latter, it is very evolved. 
Novae, which can appear as SSSs, are also generally old systems.

 Young sources can be found
only in regions containing signatures of recent star formation:
OB associations, HII regions, and SNRs. For example,
in $10^6$ years, a system traveling at
$100$ km s$^{-1}$ can travel $100$ pc, or about $27''$ at the 
distance to M31.    
If, therefore we find density enhancements of SSSs close to young stars ($\leq 1'$),  
it is very likely that some of those SSSs are young.  
It is possible that some hot white dwarfs could be young systems--
e.g., if they are PN descendants of $\sim 8\, M_\odot$ stars,
or if they accrete winds from a high-mass star.

Similarly, if some SSSs are formed close to the galaxy's center through
the tidal disruption of giants,
and if they remain supersoft for a few times $\sim 10^6$ years, then
they must be found within a few hundred pc of the nucleus.

Below we establish that there are SSSs in the bulge, disk and
halo of M31 and that their locations are such that some may be young
systems while others may be old. This study by itself
cannot, however, establish relative densities of sources in the bulge, disk and halo.
This is due to projection effects and also because  
the effects of an intervening gas
column are so powerful in hiding SSSs, lowering the 
count rate, or altering their spectra, that we need to 
couple the 
X-ray observations with detailed maps of the column density distribution.

\subsection{Bulge Sources}

Before we observed M31, it was not clear whether SSSs would be found
in galaxy bulges. High Galactic absorption prevents us from detecting
them in the bulge of the Milky Way. Predictions were difficult, not only
because of uncertainties in the models, but also because the stellar populations
near the center of galaxies like the Milky Way and M31 are complicated;
old stellar populations are expected, but there is also evidence of
recent and ongoing star formation. There were no predictions at all for QSSs.
 
{\it ROSAT} discovered only $1$ SSS in the central kpc.
{\it Chandra} finds $14$ VSSs in the central $8' \times 8'.$
Two of these are within a projected distance of $100$ pc of the
galaxy center,  and an additional $8$ VSSs lie
within a projected distance of $\sim 450$ pc of the center.

The earliest {\it Chandra} observation of M31 discovered 
an SSS within $2''$ of nucleus.   
Given the value of the SSS spatial density,  
the probability of an SSS being within a few pc of the nucleus
is very low. It therefore 
seems likely that the source is somehow related
to the presence of the nucleus. A natural explanation is that the
VSS is the hot core of a giant that was tidally stripped
by the massive central ($3 \times 10^{7} M_\odot$) BH. 
Some of the $10$ sources within $\sim 450$ pc of the nucleus could
potentially be stripped cores of giants, which have had time
to move from the galaxy center. Statistical studies of
a broad range of galaxies, as well as further work on M31,
e.g., comparing the spatial density distribution of VSSs with other
X-ray sources near the galaxy center, may help to verify or 
falsify this hypothesis. 
It is also possible that the centrally located
VSSs are interacting binaries descended from   
the young and old populations that inhabit the bulge.  

The VSSs in the bulge have higher average and median count rates
than those in the disk.
The average and median and highest count rates for the central 
field are $8.7$ ks$^{-1}$ and $2.6$ ks$^{-1}$ and $69.3$ ks$^{-1}$, respectively.
For the non-stellar sources in
Fields 1, 2 and 3, the corresponding triplets of numbers 
(average, median, and high count rates) are 
($0.69$ ks$^{-1}$, $0.35$ ks$^{-1}$, $1.38$ ks$^{-1}$),
($1.14$ ks$^{-1}$, $0.98$ ks$^{-1}$, $1.86$ ks$^{-1}$), and
($1.41$ ks$^{-1}$, $0.93$ ks$^{-1}$, $4.31$ ks$^{-1}$), respectively.
The bulge VSSs are also generally softer, as witnessed
by the fact that a larger fraction of them are SSSs identified by the
HR condition ($8$ HR sources out of $16$ bulge sources). There are
$1$, $1$, and $0$ HR source out of $5$, $5$, and $7$
 non-stellar SSSs in Fields 1, 2, and 3, respectively.

\subsection{Disk Sources}

Because of M31's tilt relative to our line of sight, it is 
difficult to identify any particular source as a disk source,
rather than a member of the halo. Nevertheless, the clustering of VSSs
in Field 2 near regions with young stars, as delineated in Figure 1 by the
presence of SNRs, indicates that some 
are likely members of the disk population.
All 5 non-stellar VSSs in Field 2 are clustered near
regions of star formation. One, 
s1-42, is in fact itself a SNR. Three of the others 
are transient and are probably X-ray binaries.
If their location near SNRs is genuine, and is not due
to projection effects, such VSSs
are likely to be young systems.

\subsection{Halo Sources} 

The VSS in Field 3 which is positioned farthest from the
major axis of the disk is the QSS n1-66. At $\sim 32'$,
or $\sim 4.8$ kpc from the major axis, and $\sim 56'$, or
almost $14$ kpc, from the nucleus, this source is likely to be in the halo
of M31.
Other non-stellar sources, including the QSSs n1-46 and n1-31, are also
far enough away from the disk to be members of halo.
Some VSSs in Field 1 are as far as $\sim 80'$ ($18$ kpc) from the nucleus,
and are also likely to be halo sources. These include the QSS s2-10,
which is a good candidate for a symbiotic system. Its location in the halo
is consistent with the fact that symbiotics
can be members of old populations. 

\section{Quasisoft Sources}

In order not to miss some of the
highest-T ``classical'' SSSs, those most likely to be high-mass
accreting white dwarfs, we had to assume the risk of selecting sources of
even higher temperature than we had originally wanted.
In figure 4 we present examples of the ranges of temperatures and luminosities
of sources selected by our algorithm.
Because we had too few counts to fit spectra, we used a comparison
of the data, binned into our $3$ broad spectral bands,  with models.
Count rates for the models had been computed using the PIMMS software
(Mukai 1993, http://asc.harvard.edu/toolkit/pimms.jsp).
We attempted to match the total counts and distribution of
counts in $S,$ $M,$ and $H,$
with models that had been computed by PIMMS. Each model is a 
black body characterized by a temperature, luminosity, and value 
of $N_H$ (The PIMMS models should be
roughly consistent with the detector sensitivity of AO2.)
If the count rates in each band agreed with the model (within
the $1\, \sigma$ uncertainty limits determined by Poisson statistics), 
we considered the model to 
be a possible match.

We considered all $33$ non-stellar sources--i.e., the
sources we think are most likely to be associated with 
M31 itself, although a few could be background objects.
We sorted the sources according to the step of our
algorithm which selected them.

 Each match is represented by a symbol in the figure.
The bottom curve corresponds to $N_H=1.6 \times 10^{21}$ cm$^{-2}$.
We chose this value because it is typical of values derived from
spectral fits of M31's XRSs. 
A unique symbol is used for each source for which a match was found.
Clear differences can be seen among the categories. For example,
SSS-HRs are mostly compatible with WD models, whereas
QSS-FNOH's are mostly not. To test whether increased absorption 
can cause the QSSs to be associated with softer models,
we also considered $N_H=6.4 \times 10^{21}$ cm$^{-2}$.    
Although for some QSSs the softening was significant, for
others it was not, or else produced intrinsic luminosities
much higher than expected. 

The hardest of the quasisoft sources could be neutron stars
with little hard emission (or perhaps beamed hard emission)
and unusually large photospheric radii (Kylafis \& Xilouris 1993).  
They could be white dwarfs as well; in this case, the emission we observe 
is harder than
expected, even for white dwarfs with masses near the Chandrasekhar mass.
This could indicate that the emission is emanating from only a 
limited portion of the white dwarf surface,
or possibly that there has been upscattering of 
photons that originated from the 
white dwarf.
Given the rough range of luminosities and temperatures that characterize QSSs,
it is likely that more than one physical model realized in nature can cause
an X-ray source to have the properties of a QSS. 
The key point is that, 
whatever physical model or models apply, QSSs appear to represent states  
that have not yet been well studied. In fact there are no known examples
of these sources in the Galaxy or Magellanic Clouds.
Perhaps this is not surprising, given the survey data available
at present. For example, The {\it ROSAT All-Sky Survey} ({\it RASS}) would have
detected QSSs in the Galaxy, but would not have found them to be
particularly soft, since the sensitivity limit of the PSPC
was just slightly harder ($\sim 2.5$ keV) than the emission from QSSs (See Figure 2).
The {\it Rossi X-Ray Timing Explorer} ({\it RXTE}) would not have detected
QSSs at all, since it is not sensitive to photons with energies below $2$ keV.
Approaches to identify QSSs have been discussed in \rd\ et al.\, 2004b)  

We mention one intriguing possibility: that some QSSs could be accreting
BHs of intermediate  mass (IMBHs). 
A natural explanation is provided by an optically thick, geometrically thin 
disk around an accreting black hole.
If we identify the source temperature with the temperature of the inner disk,
located at the radius of the last stable orbit, and assume that $10\%$ 
of the accretion energy is emitted, then the mass of the
BH accretor is given by 
\begin{equation}
M_{BH}\sim 80\, M_\odot\ \Big[{{200\, {\rm eV}}\over{k\, T}}\Big]^2\
                \Big[{{L}\over{10^{38} {\rm erg/s}}}\Big]^{{1}\over{2}}, 
\end{equation}
up to a factor which takes accretion efficiency, exact location
of the disk's inner edge, spectral hardening, and disk orientation
into account.  
This example illustrates that some of the quasisoft models could be
accreting intermediate-mass BHs.
It is important to note that even the softest sources could be 
accreting intermediate-mass BHs. 
Figure 4 simply illustrates that there are some soft sources
in M31 that may be inconsistent with white dwarf models;
these are then the most obvious candidates for other models, 
such as the BH model.

\section{Luminosity Distributions}

In each field, there are VSSs with count rates below $1$ ks$^{-1}$.
In Field 1, the $2$ lowest count-rate sources have  
count rates of $0.27$ ks$^{-1}$ and
$0.35$ ks$^{-1}$ . In each of Fields 2 and 3 there are 
$3$ sources with count rates ranging from 
$0.22$ or $0.34$ ks$^{-1}$ to $\sim 1$ ks$^{-1}$.
In Field 2, there are $5$ sources with count rates below
$1$ ks$^{-1}$. 

Because the luminosity, temperature, and column density each
play a crucial role in determining the count rates of SSSs,
the low-count-rate sources undoubtedly represent a range of
source luminosities.    
For the AO2 observations, the range of VSS luminosities
of the low-count-rate sources was between $\sim 10^{35}$ erg s$^{-1}$
(for a 100 eV source behind a column of $7 \times 10^{20}$ cm$^{-2}$),
to just over $10^{36}$ erg s$^{-1}$
(for a $50$ eV source behind a column of $1.5 \times 10^{21}$ cm$^{-2}$).

Although we cannot determine which among our sources have the
lowest luminosities, we can say that we
are likely detecting sources with luminosities
below $10^{37}$ erg s$^{-1}$. The significance of this 
number is that it represents the luminosity of an $0.6\, M_\odot$
 white dwarf that is burning
hydrogen to helium in a quasi-steady manner. 
We conjecture that at least some of the lower luminosity
sources may correspond to nuclear-burning
 white dwarfs of lower mass or bright CVs
(Greiner et al. 1999, Greiner \& \rd\ 1999, Patterson et al. 1998).

\section{Conclusions}
Our work with M31 demonstrates the challenges and rewards of
studying SSSs in other galaxies. The primary challenge is posed
by the low count rates, making it difficult to select sources
based on spectral criteria.
For SSSs the situation is  complicated because
(1) we don't know {\it a priori} whether one or more physical models apply,
and (2) for any specific physical model,  we cannot uniquely predict the
spectrum from the accretor, disk, corona, interactions with the ISM, etc.
We have therefore chosen to develop and apply an algorithm for source selection
that
has been tested on simulated and real data from other galaxies. 
\footnote{ The other galaxies were chosen because there is little absorption
along the line of sight to them and in some cases (e.g., M101, M83),
the orientation is more face-on than M31's, hence is favorable for the
detection of VSSs.} 
(See \rd\ \& Kong 2003b, c.) 
The algorithm categorizes sources according to the details of
how they were selected, allowing us to select those most likely
to be intrinsically soft. Future investigations can now focus on
any subset of the VSSs identified by our algorithm.

In M31 we find 33 VSSs that are not foreground stars. Five of these are 
in Field 1, five are in Field 2, seven are in Field 3,
and sixteen are in the central
field. Comparative studies of the {\it ChaMP} fields to estimate the numbers of 
background VSSs indicate that all of these sources are likely to 
be in M31.
Two of the sources (r2-56 and s1-42) appear to be associated with SNRs.
One of these, r2-56, is resolved at X-ray, optical, and radio wavelengths.
This SNR exhibits no emission above $1.1$ keV and was selected by the HR
conditions, the strongest selection criteria. s1-42 satisfies the $3\sigma$ 
criteria, the second strongest set of selection criteria.

The sixteen bulge VSSs include the brightest, softest VSSs in M31.
The bulge sources are highly concentrated within a projected distance of
 $450$ pc from the galaxy center. 
The surface density and count rates of VSSs in the disk field are smaller,
and the sources are, on average, harder than those in the bulge. 
It remains to be seen if the differences between the disk and bulge are due
to greater absorption in the disk or if they are intrinsic to the sources. 

M31's VSSs are highly variable. Thirteen out of 15 (non-SNR) bulge
VSSs rose above or fell below the {\it Chandra} and {\it XMM}
detectability limits on time scales of months, as did 3 of 5 non-stellar
SSSs in Field 2. There have been fewer checks for variability in 
Fields 1 and 3, as they were not covered by {\it XMM}. 
Nevertheless, 8 of the 8 {\it ROSAT} SSSs scattered across 
Fields 1, 2 and 3 that could be checked for variability had faded
below detectability between 1991 and 2000/2001. The fading of the SSSs
that have fallen below the detectability limits cannot be explained 
by assuming that the SSSs are novae. 
The required rate of novae would be too high to be compatible with other data
(Shafter \& Irby 2001).
Furthermore we find sources turning on as well as turning off, with 
on-off-on or off-on-off behavior observed for some VSSs. 
The most viable conjecture is that 
the variable VSSs are X-ray binaries. 

Disk VSSs in Field 2 appear to be clustered near star-forming regions, 
indicating that they may be young systems. Others appear to be in the halo
of M31 and may be old.
The results match what we have seen in other galaxies. In M101, e.g., 
which we view face-on, VSSs are predominantly found in the spiral arms,
(\rd\ \& Kong 2003a, c). In M104 (the Sombrero), on the other hand, which we view edge-on,
some VSSs are clearly located several kpc away from the disk. 

In the center and disk we find VSSs with count rates between $0.25$ and $1$ 
counts $ks^{-1}$. These are likely to have luminosities between 
$10^{35}-10^{37}$erg$s^{-1}$, lower than those of the Magellanic Cloud and 
Galactic SSSs that defined the class through {\it ROSAT} observations
in the early 1990's. Some of these sources may be bright CVs with 
relatively high accretion rates (Greiner and \rd\ 1999, Greiner et al. 1999,
Patterson et al. 1998). 

Considering all of the fields we studied, 
10 SSSs were discovered by the HR condition and 5 by the next most selective
criteria, the $3\sigma$ conditions.
Eighteen non-stellar sources satisfy conditions that are somewhat less strict,
and we identify these as QSSs 
Thirteen of 
these exhibit no significant emission above 2 keV ]
and 5 allow some emission in the H band, as long as the 
fraction of such photons is small. 

The regions containing the largest numbers of detectable VSSs 
are the four $8' \times 8'$ squares covered by S3. Even in each of 
Fields 1 and 3, far from the nucleus, there are $3$ non-stellar 
VSSs in S3. In regions close to but not overlapping the nucleus,
the numbers of sources in an $8' \times 8'$ field should lie between $3$ and $16$. 
If we consider that there are $>$ forty $8' \times 8'$
fields within $D_{25}$, and assume that each houses four VSSs, 
an A0-2 S3 survey of M31 would have detected $>300$ VSSs. 
To derive the total underlying population, absorption effects 
need to be taken into account; this will be done separately.

In terms of the properties and possible natures and origins, 
the VSSs of M31 are a more diverse group than we had anticipated.
{\it HST} ACS observations to identify dim optical counterparts, 
{\it XMM} data on the disk fields, and {\it Chandra} observations 
of the region  will help us to unravel the mystery of 
their fundamental natures. 

\begin{acknowledgements}
This work was supported by NASA under an LTSA grant,
NAG5-10705. A.K.H.K. acknowledges support from the Croucher Foundation.
R.D. would like to thank P. Green \& P. Plucinsky for discussions and 
R. Remillard for providing access to an unpublished paper.
We would also like to thank the referee, P.A. Charles, for
comments that have helped to improve the paper.  
\end{acknowledgements}

\end{document}